\documentclass[%
 reprint,
 amsmath,amssymb,
 aps,showpacs,prl
]{revtex4-1}

\usepackage{graphicx}
\usepackage{dcolumn}
\usepackage{bm}
\newcommand{\bb}{\bibitem}

\begin{document}

\preprint{APS/123-QED}

\title{Generic anisotropic Lifshitz scalar field theory: masslesslike  massive minimal subtraction}
\author{Marcelo M. Leite}%
 \email{marcelo.mleite@ufpe.br}
\affiliation{Laborat\'orio de F\'\i sica Te\'orica e Computacional, Departamento de F\'\i sica,\\ Universidade Federal de Pernambuco,\\
50670-901, Recife, PE, Brazil}


\begin{abstract}
 {\it We formulate the simplest minimal subtraction version for massive $\lambda \phi^4$ scalar fields with $O(N)$ symmetry for generic anisotropic Lifshitz spacetimes.  We introduce the simplest geometric concepts to define it as a special manifold. Then we restrict ourselves to flat Euclidean spaces.  An appropriate 
partial$-p$ operation is applied in the bare two-point vertex function diagrams, which 
separates the original diagram into a sum of two different integrals which are the coefficients 
of the corresponding polynomials in the mass and external momentum. Within the proposed method, the coefficient of the mass terms can be discarded and we obtain  a minimal subtraction method  almost identical to the same scheme in the massless theory in {\it every external momentum/mass subspace}. We restrict our demonstration of the method up to three-loop order in the two-point vertex part.  We verify its consistency through a diagrammatic computation of static critical exponents, which validates the universality hypothesis.}         
\end{abstract}

\maketitle

\par Analytical methods in quantum field theory are rather compelling for at least three reasons. First, they provide a simple physical interpretation of the result. Second, a large number of intermediate steps might cancel out among each other producing a null effect. Third, we can figure out new insights that would be impossible otherwise using numerical methods. Specifically, static anisotropic $m$-axial Lifshitz critical behaviors \cite{RLS} were formulated as $\lambda \phi^{4}$ field theory, but their actual exact analytical solution in perturbation theory is still lacking. 
\par The orthogonal approximation devised for this problem either in its 
massless \cite{L1,L2} or massive \cite{CL1} settings yields a systematic analytical solution to all orders in perturbation theory in momentum space. "Lifshitz space" \cite{CL1} and "generalized anisotropic Lifshitz spaces" were defined \cite{CL2}. This approximation in the massive framework to anisotropic $m$-axial Lifshitz points suggested naturally the concept of a "partial-$p$" operation defined in Lifshitz space(time)s from a particular Lifshitz scalar field theory using the $BPHZ$ method \cite{SCL}. Nevertheless, a decisive explicit proof of how Lifshitz spacetimes come up in this context is still lacking. Can we give a truly spacetime description of Lifshitz type scalar quantum fields? Those theories are renormalizable. Can we find a simpler minimal subtraction method in order to achieve a maximal simplification in the renormalized theory?
\par In this Letter we discuss how generic anisotropic Lifshitz scalar quantum field theory 
($QFT$) can be defined in a Eucliedean flat space as a particular case from a metric tensor representing a special curved higher dimensional spacetime.  We then introduce a partial-$p$ operation with many mass scales. We apply it in a particular  type of one-particle  irreducible ($1PI$) formalism and obtain a "masslesslike" massive minimal subtraction renormalization scheme, similarly to the recent method developed for pure $\phi^{4}$ theory \cite{LL}. This simplified framework does not contain tadpole insertions in any $1PI$ vertex parts, therefore requiring a minimal number of diagrams.  We discuss how the $m_{n}$ subspaces can be renormalized independently. We compute critical exponents diagrammatically that checks the consistency of our method. Whenever $m_{n}=0$ for $n=2,...,L$ {\it all} the results here reduce to those from \cite{LL}.  
\par Consider a bare scalar field Lagrangian density (Euclidean version) with $O(N)$ symmetry defined for the most general anisotropic Lifshitz static multicritical behavior ($CECI$ model) whose action(/Helmholtz free energy)  is:
\begin{eqnarray}\label{1}
&S= \int \Pi_{n=1}^{n=L} d^{m_{n}} x_{n} \Bigl[\frac{1}{2}
|\bigtriangledown_{(d- \sum_{n=2}^{L} m_{n})} \phi\,|^{2} +
\sum_{n=2}^{L} \frac{\sigma_{n}}{2}\nonumber\\
&\times |\bigtriangledown_{m_{n}}^{n} \phi\,|^{2} + \sum_{n=2}^{L} \delta_{0n}  \frac{1}{2}
|\bigtriangledown_{m_{n}} \phi\,|^{2}
+ \sum_{n=3}^{L-1} \sum_{n'=2}^{n-1}\frac{1}{2} \nonumber\\
&\times \tau_{nn'} |\bigtriangledown_{m_{n}}^{n'} \phi\,|^{2} + \frac{1}{2} \mu_{0}^{2n} \phi^{2} + \frac{1}{4!}\lambda_{0n} (\phi^{2})^{2}\Bigr] .
\end{eqnarray}
\par We particularize our discussion throughout to the cases $\delta_{0n} = \tau_{n n'} =0$ . There are independent subspaces $(m_{1},...,m_{L})$ (with coordinates ($x_{(1)}^{i_{1}}, ...,x_{(L)}^{i_{L}}$) with $i_{1}= 1,..,m_{1},...,i_{L}=1,...m_{L}$, respectively, such that $d=m_{1}+...+m_{L}$ is the space dimension . The above action has a purely quantum nature (for details see Ref. \cite{L2}) and recall that in statistical mechanics the path integrals are weighted by $e^{-S}$.    
\par In the terminology of  Lifshitz quantum field theory the power of momentum is named $z$. In theory of membranes in quantum criticality, world-volume scalars (space vectors) are  embedded in $26$-dimensional target space with additional Lifshitz matter composed of extra terms with world-volume scalars presenting second derivatives  (irrelevant deformations) \cite{Horava1}. The results for the classical theory of gravity in this setting is that in $z=1$ it is very close to general relativity (GR) in the infrared ($IR$). For $z=2$, its behavior is restricted to the ultraviolet ($UV$). It presents an extra propagating degree of freedom in the ($UV$), although holography can also be formulated \cite{Horava2}. The values $z=3,4$ were also studied in this gravity model.  
\par Another approach to delve further into the properties of the dimension of a geometric object
$\mathcal{M}$ is the utilization of the definition of its spectral dimension, namely, the effective dimension of the object as seen by an appropriate diffusion process (or random walker) occurring on it.  Among other results, like reproducing the space dimension of the flat case $\mathcal{M}= \mathcal{R}^{d}$, the transition from $z=1$ to $z=3$ at fixed space dimension was shown to change the spectral dimension of the universe from short- to long-distance regimes \cite{Horava1}. This argument agreed with simulations of lattice quantum gravity via causal dynamical triangulations \cite{Amb}.
\par In the works \cite{Horava1,Horava2} the 
irrelevant deformations introduced are always massless. In particle physics, the inclusion of Lifshitz terms with changing values of $z$ for massless scalars, spinors and gauge fields \cite{Kawamura} addresses the proton stability in the standard model ($SM$), whereas 
by adding a higher derivative scalar field of the gauge group \cite {Kaneta}
explains the increase in the fermion masses in the $SM$ . Unfortunately, only tree-level effects were discussed in \cite{Kawamura,Kaneta}. The mass of a scalar field can be relevant to change  the scaling dimension of some coordinates when it interacts with charged Lifshitz black branes/holes \cite{Z}, although there is some controversy about the existence of slowly rotating black holes in Horava's gravity \cite{Enrico}. The  aftermath of Horava's proposal has been developed in a number of papers on holography \cite{Nishioka}, black holes \cite{Cai} and cosmology \cite{Wang}. In addition, the study of Lifshitz field theory for scalars and gauge fields 
under special conditions was utilized to explain the time delays in gamma-ray bursts \cite{Chen}.  
\par In our considerations of Lifshitz space-times here we have a more modest aim. Consider the effective Lagrangian of a gravitational field in $d=\sum_{n=1}^{L} m_{n}$ coupled with a self-interacting quantum scalar field $\phi$ whose action is given by
\begin{eqnarray} \label{2}
&S_{QFT}=\int \Pi_{n=1}^{n=L} d^{m_{n}} x_{n} \sqrt{-g} \Bigl[\frac{1}{2} \partial_{M} \phi \odot g^{MN}  
 \odot \partial_{N} \phi \nonumber\\
&+ \frac{\mu_{0 n}}{2} \phi^{2} + \frac{\lambda_{0 n}}{4!} (\phi^{2})^{2} \Bigr].
\end{eqnarray}
\par The special product reflects the special structure of the manifold, whose metric tensor $g^{MN}$ depends on derivatives in the competing subspaces. Furthermore, the determinant of the metric tensor $g_{MN}$ can also be defined.  The discussion of these issues would take us too far afield in the present work and we relegate these details 
to a future publication. Nevertheless, we will explain the most basic aspects of the manifold structure which originates such features. For the time being is sufficient to say that for flat subspaces ${m_{n}}$ of Euclidean signature, it is possible to prove that Eq. (\ref{1}) is retrieved from Eq.(\ref{2}) for the appropriate choice of the metric tensor with the prescriptions outlined above.
\par The subspace $m_{n}$ contains $p_{n}^{2n}$ in the bare propagator. We can set 
$\sigma_{n}=1$ \cite{L2}in Eq.(\ref{1}), but this affects the canonical dimensions of the coordinates in different subspaces.  Indeed, if  $[x_{1}^{i_{1}}] = M^{-1}$ (or $[p_{1}^{i_{1}}]=M)$, in the $m_{n}$ subspace one has instead $[x_{n}^{i_{n}}] = M^{-\frac{1}{n}}$ (or 
$[p_{n}^{i_{n}}]= M^{\frac{1}{n}}$). This dimensional redefiniton produces an interesting effect: the "dilution" of the dimensions contained in the competing subspaces.
\par We are now at a position to give a set-theoretic interpretation of the dimension in the competing subspaces.  We use the definition given by Kolmogorov for the "box-counting dimension" of a set of points.  (It gives the same result for the calculation of the dimension as that using the Hausdorff's method.  Kolmogorov's method is, however, much simpler. )The definition is $D_{0}= \frac{logN(\delta)}{log(\frac{1}{\delta})}$ in the limit $\delta \rightarrow 0$ \cite{Mon}. Here $N(\delta)$ is the maximal number of identical "boxes" with side $\delta$ needed to cover the entire set of points. The rigorous definition of the usual ternary Cantor set is in  terms of the closed compact interval $[0,1]$ of the real line.  Since the interval $[0,1]$ has a one-to-one correspondence to the ternary Cantor set and inherits its topology from the real line,  the same topology is induced on the ternary Cantor set.  We shall relax the compact interval in our generalized Cantor sets to be defined below.  Loosely speaking, the same topology of the real line is induced on each of them. 
\par Le us apply Kolmogorov's definition of the dimension to compute this object for all of them.   We exemplify the "dilution"  in the uniaxial case for $m_{2}$ with a generalized Cantor set, which is initially a line of length $l$.  
In the first iteration, it is divided into 4 pieces of equal length 
$\frac{l}{4}$ and we keep only the two disconnected intervals $[0,\frac{l}{4}] \bigcup [\frac{3l}{4},l]$. Iterating $p$ times using the same steps we conclude that $N= 2^{p}$ whereas $\delta=\frac{l}{2^{2p}}$ which leads to $D_{02}= \frac{1}{2}$ in the limit $p \rightarrow \infty$.  Let us call $F_{2 1}$ the set of intervals obtained in the first interaction, $F_{2 2}$ the set of intervals in the second interaction and so on.  We can associate this subspace with the generalized Cantor set $L_{2}$, defined by $L_{2} = \bigcap_{n=1}^{\infty} F_{2 n}$.  To see that this set has zero length,  note that the total length withdrawn ($lw_{2}$) from the interval$[0,l]$ is given by the expression 
$lw_{2}= l \sum_{p=1}^{\infty}\frac{1}{2^{p}}=l$, therefore proving our assertion. 
\par  Consider $m_{3}$. In the first iteration, divide the interval by 27, and choose only the intervals 
$F_{3 1}= [\frac{8l}{27} ,\frac{l}{3}] \bigcup [\frac{17l}{27}, \frac{2l}{3}] \bigcup [\frac{26l}{27},l]$ and so forth, analogously to our above observations for the $m_{2}$ subspace. In the 
$p$-th iteration $N= 3^{p}$ and $\delta= \frac{l}{3^{3p}}$. We obtain 
$D_{03}= \frac{1}{3}$ in the limit $p\rightarrow \infty$.  The associated generalized Cantor set to this subspace is 
 $L_{3} = \bigcap_{n=1}^{\infty} F_{3 n}$.  Note that the total length withdrawn of the interval of length $l$  now reads 
 $lw_{3}= 8l \sum_{p=1}^{\infty}\frac{1}{9^{p}}=l$ which implies that $L_{3}$ also has zero length. 
 \par This can also be done for $m_{n}$ yielding $D_{0n} = \frac{1}{n}$, $n=2,...,L$, whose associated generalized Cantor space is $L_{n} = \bigcap_{p=1}^{\infty} F_{n p}$.  In this case, $lw_{n}=l (n^{n-1} -1) 
 \sum_{p=1}^{\infty}\frac{1}{n^{(n-1)p}}=l$ and $L_{n}$ has zero length as well (or zero Lebesgue measure).  
Note that Smith-Volterra-Cantor sets have positive Lebesgue measure.  Since the new generalized Cantor sets above introduced share the zero Lebesgue measure property with the original ternary Cantor set but never reduce to it, we find appropriate to name them "Cantor-Leite" sets $L_{n}$.
\par Although the dimension of the different subspaces computed with our method gives basically the same result from Ref. \cite{Horava1}, the latter did not associate this with a special set different from $R^{m_{n}}$,  perhaps because the topology of the various generalized Cantor-Leite sets above described inherit the topology of  $R^{m_{n}}$ . However, in that work there are no extra dimensions: although this possibility was slightly mentioned there, all spatial dimensions belong to the $m_{n}$ subspace ($n=2,3,4$). 
\par Whenever $n^{np} \gg l$, the result consists of several disconnected continuous pieces but  has fractal dimension. In this simple model,  it can be interpreted as the attractive and repulsive combined effects of gravity in the extra dimensions in a close analogy with the $CECI$ model in statistical mechanics .  This competition "dilutes" them so to speak; see below.  Even though every Cantor-Leite set inherits the topology of the real line, its local structure demands a different metric structure  from that of $\mathcal{R}$.  Let $L_{n}^{m_{n}}$ be the set representing the $m_{n}$ competing subspace.  We can perform integrals as if we were working in $\mathcal{R}^{m_{n}}$, but clearly additional geometric structure is necessary since $K^{m_{n}}$  is  naturally different from $\mathcal{R}^{m_{n}}$.  Some care must be exercised.  All generalized Cantor sets discussed above $L_{n}$ were shown to have zero length.  This does not mean that  $K_{n}$ coincides with the empty set, since certainly several intervals were left in its construction.  So the integrals we referred to above should be computed with the Riemann measure and the integrals over the real lines are well-defined.  Thus the manifold corresponding to the extended space has the form $\mathcal{M}^{d} = \mathcal{M}^{1, m_{1}-1} \times L_{2}^{m_{2}} 
\times L_{3}^{m_{3}}...\times L_{L}^{m_{L}}$. 
\par Considering just flat competing subspaces, a realization of this spacetime manifold as a nontrivial metric space 
can be defined through its metric tensor whose components are generalizations of $g_{\mu_{1} \nu_{1}} (x^{\rho_{1}})$ of the curved metric on the subspace $m_{1}$.  
\par With a slight change of notation in the coordinates for the time being, the metric tensor defined on the $m_{n}$-dimensional subspace has the form
\begin{eqnarray}\label{3}
&  \hat{g}_{i_{n} i'_{n}}  = \delta_{i_{n} i'_{n}} (-1)^{n-1} \frac{ \tilde{\sigma}_{n}^{-1}}{m_{n}} \sum_{i_{r},i_{s} =1}^{m_{n}} 
(y^{i_{r}} y^{i_{s}})^{n-1}[ \frac{1}{(2n-2)!} \delta_{i_{r} i_{s}}  \nonumber\\
&\;\; +\frac{1}{(n-1)^{2}}(1 - \delta_{i_{r} i_{s}})]\hat{i}_{r} \hat{i}_{s}  .
\end{eqnarray}
\par This metric has a double tensor structure, one coming from the Kronecker delta and the other coming from the $m_{n}^{2}$ dyadics components manifested by the products  $\hat{i}_{r} \hat{i}_{s}$ ($\hat{i}_{r}. \hat{i}_{s}=\delta_{rs}$).  Another important point is the appearance of the factors $(-1)^{n-1}$ in the metric pertaining to the $m_{n}$ competing subspace. They were introduced so that the signs of all higher derivative terms after integral by parts coincide with those from Eq. (\ref{1}) when surface terms are discarded whenever we perform the identification $\tilde{\sigma}_{n} \equiv \sigma_{n}$.
\par Within the $m_{1}$ subspace the metric does not have to be diagonal.  Its inverse for a curved space $g^{\mu_{1} \nu_{1}}$ satisfies 
$g_{\mu_{1} \nu_{1}}g^{\nu_{1} \mu_{2}}= g^{\nu_{1} \mu_{2}} g_{\mu_{1} \nu_{1}} = \delta_{\mu_{1}}^{\mu_{2}}$.  It is worth emphasizing  that the different subspaces do not mix which other.  We also have to be careful to contract the dyadics in order to obtain a pure number.  
\par The statistical mechanics model defined by Eq. (\ref{1}) has a purely quantum nature (the exchange interactions, with competing effects due to different signs among farther neighbors in a generalized Ising model). In our $QFT$ of a self interacting scalar field minimally coupled to a gravitation field in a higher dimensional space,we can think of the competing subspaces as resulting from gravitational and antigravitational interactions in extra dimensions ("dark matter" effects \cite{Sengupta}). Thus something similar to quantum operators must arise. 
\par This quantum structure of the extra dimensions can be partially implemented through the inverse metric.  In order to achieve this, define $\hat{g}^{i_{n} i'_{n}}$ as
\begin{eqnarray}\label{4}
&  \hat{g}^{i_{n} i'_{n}}  = \delta^{i_{n} i'_{n}} (-1)^{n-1}\frac{ \tilde{\sigma}_{n}}{m_{n}} \sum_{i_{r},i_{s} =1}^{m_{n}} 
(\frac{\partial}{\partial y^{i_{r}}} \frac{\partial}{\partial y^{i_{s}}})^{n-1}  [\delta_{i_{r} i_{s}}  \nonumber\\
&\;\; + (1 - \delta_{i_{r} i_{s}}) ]\hat{i}_{r} \hat{i}_{s}  .
\end{eqnarray}
Therefore, the matrix multiplication in ordinary spacetime turns into a new structure: the usual matrix multiplication in the $m_{1}$ subspace whereas this matrix multiplication includes the double internal product of the independent dyadics sector in the competing subspaces $m_{n}$ ($n=2,...L$).  Consequently this definition yields the identity $g^{MP} . .  g_{PN}=\delta_{N}^{M}$.
\par In first quantized 
quantum mechanics, a classical quantity takes operator values. Our description corresponds to the semiclassical (first quantized theory) of the metric tensor along the extra directions (competing subspaces $m_{n}, n=2,...,L$).  The subtle point is that  only the contravariant metric tensor in the competing subspaces implements this idea.   
It is not surprising, since the origin of the Lagrangian (\ref{1}) is purely quantum (the exchange couplings have no classical analogue). The (leading quantum) effect of attractive/repulsive competition provoked by the semi-classical gravitational field {\it at short distances ($UV$)} is the appearance of the competing subspaces. 
\par The Levi-Civita connection $\Gamma^{M}_{NP}$ vanishes for arbitrary 
$M, N, P\neq \mu_{1},\nu_{1},\rho_{1}$.  Hereafter, our choice of flat Euclidean metric in the $m_{1}$ subspace implies 
$\Gamma^{M}_{NP}=0$.
\par The masses are inversely proportional to the many independent correlation lengths $\xi_{n}$ in this multicritical behavior. (Henceforth, the Euclidean version of the metric will interest us.) We choose to keep them different to tackle the problem purely from the field theory viewpoint \cite{CL2}. Anticipating future applications in the high energy (UV) regime, we discuss its renormalizability. Note that 
$\epsilon_{L} = 4 + \sum_{n=2}^{L} \frac{(n-1)}{n} m_{n} - d$, $m_{1}$ varies but the value of $m_{n}$ ($n \neq 1$) will be kept fixed throughout. 
\par  For the time being we consider only the primitively divergent vertex parts without worrying about graphs with tadpole insertions. How we achieve this, however, will be explained later during our explicit discussion of these vertex parts. 
\par We begin with the integrals up to two-loop order in the $1PI$ four-point 
$\Gamma^{(4)}$ (and composed field $\Gamma^{(2,1)}$ since they are related with each other). Although not needed in what follows we write them with arbitrary external momenta in all subspaces in terms of a reference mass $\mu_{n}^{*}$. The mass 
$\mu_{n}$ is then interpreted as the restriction of $\mu_{n}^{*}$ when we specify the particular subspace with nonvanishing external momenta, thanks to the orthogonal approximation. They read:
\begin{subequations}\label{5}  
\begin{eqnarray}
&I_{2}(P_{(n)}, \mu_{n}^{*}) = \int \frac{\Pi_{n=1}^{L} d^{m_{n}} q_{(n)}}{((\sum_{n=1}^{L} q_{(n)}^{2n}) + \mu_{n}^{* 2n})}\nonumber\\
& \times \frac{1}{[(\sum_{n=1}^{L} (q_{n)}+P_{(n)})^{2n}) + \mu_{n}^{* 2n}]},\label{5a}\\
&I_{4} (p_{i (n)}, \mu_{n}^{*}) = \int \frac{\Pi_{n=1}^{L} d^{m_{n}} q_{1(n)} d^{m_{n}} q_{2(n)}}
{[(\sum_{n=1}^{L} q_{1(n)}^{2n}) + \mu_{n}^{* 2n}][(\sum_{n=1}^{L} q_{2(n)}^{2n}) + \mu_{n}^{* 2n}]} \times 
\nonumber\\
& \frac{1}{[(\sum_{n=1}^{L}(P_{(n)}-q_{1(n)})^{2n}) + \mu_{n}^{* 2n}]}\nonumber\\
& \times \frac{1}{[(\sum_{n=1}^{L}(q_{1(n)} - q_{2(n)} + p_{3(n)})^{2n}) + \mu_{n}^{* 2n}]}, \label{5b}
\end{eqnarray}
\end{subequations}
where $P_{(n)}= (p_{1(n)} + p_{2(n)}, p_{1(n)} + p_{3(n)}, p_{2(n)} + p_{3(n)})$ are external momenta in arbitrary subspaces. These integrals will  not be of particular interest within the context of the present method . We just write down their  $\epsilon_{L}$-expansions as
$I_{2} (p_{(n)},\mu_{n}^{*})=\frac{\mu_{n}^{* - n \epsilon_{L}}}{\epsilon_{L}} \Bigl[1 + (h_{m_{L}}-1) \epsilon_{L} -\frac{\epsilon_{L}}{2} L(P_{(n)},\mu_{n}^{*})\Bigr]$ and $I_{4} =\frac{\mu_{n}^{* - 2n \epsilon_{L}}}{2 \epsilon_{L}^{2}} \Bigl[ 1 + (2h_{m_{L}} - \frac {3}{2})\epsilon_{L} 
- \epsilon_{L} L(P_{(n)},\mu_{n}^{*})  \Bigr]$, where $h_{m_{L}}= 1- \frac{1}{2}(\psi(1) - \psi(2 - \sum_{n=2}^{L} \frac{m_{n}}{2n}))$ and 
 $L(P(n),\mu_{n}^{*}) = \int_{0}^{1} dx ln\Bigl[\frac{\sum_{n=1}^{L} P_{(n)}^{2n}}{\mu_{n}^{* 2n}} x(1-x) + 1 \Bigr]$. The conventions are detailed in Refs. \cite{L2,CL2}.     
\par By the same token, consider the two- and three-loop contributions of the two-point vertex parts with arbitrary external momenta $p_{(n)}$ in arbitrary subspaces $m_{n}$ with the reference mass $\mu_{n}^{*}$. Their manipulation is the key ingredient in our method. Their  Feynman integrals $I_{3}$ and $I_{5}$, respectively, can be written as
\begin{subequations}\label{6}
\begin{eqnarray}
&I_{3}(p_{(n)},\mu_{n}^{*}) = \int \frac{\Pi_{n=1}^{L} d^{m_{n}} q_{1(n)} d^{m_{n}} q_{2(n)}}
{[(\sum_{n=1}^{L} q_{1(n)}^{2n}) + \mu_{n}^{* 2n}][(\sum_{n=1}^{L} q_{2(n)}^{2n}) + \mu_{n}^{* 2n}]}\nonumber\\
&\frac{1}{[(\sum_{n=1}^{L}(q_{1(n)} + q_{2(n)} + p_{(n)})^{2n}) + \mu_{n}^{* 2n}]},\label{6a}\\
&I_{5}(p_{(n)},\mu_{n}^{ *}) = \int \frac{\Pi_{n=1}^{L} d^{m_{n}} q_{1(n)} d^{m_{n}} q_{2(n)} 
d^ {m_{n}} q_{3(n)}}{[(\sum_{n=1}^{L} q_{1(n)}^{2n}) + \mu_{n}^{* 2n}][(\sum_{n=1}^{L}q_{2(n)}^{2n}) + \mu_{n}^{* 2n}]}\nonumber\\
&\frac{1}{[(\sum_{n=1}^{L} q_{3(n))}^{2n}) + \mu_{n}^{* 2n}][(\sum_{n=1}^{L}(q_{1(n)} + q_{2(n)} + p_{(n)})^{2n}) + \mu_{n}^{* 2n}]}\nonumber\\
& \times \; \frac{1}{[(\sum_{n=1}^{L}(q_{1(n)} + q_{3(n)} + p_{(n)})^{2n}) + \mu_{n}^{* 2n}]},\label{6b}
\end{eqnarray}
\end{subequations}
\par The "partial-$p$" operation \cite{tHV} appropriate to our aim is written as 
\begin{equation}\label{7}
1 = \frac{1}{d_{eff}} \Bigl(\frac{\partial q_{1}^{m_{1}}}{\partial q_{1}^{m_{1}}}
+\sum_{n=2}^{L} \frac{1}{n} \frac{\partial q_{1(n)}^{m_{n}}}{\partial q_{1(n)}^{m_{n}}} \Bigr),
\end{equation} 
where $q_{i(n)}$ is each loop momentum of the $m_{n}$ subspace. This operation is applied to each loop diagram (the summation convention over repeated indices is implicit). Here 
$d_{eff} = d - \sum_{n=2}^{L} \frac{(n-1)}{n} m_{n}= 4 - \epsilon_{L}$. Note that no matter how many competing subspaces permitted by the problem, the effective dimension is always related to the critical dimension in the same way as in the usual $\phi^{4}$ with $\epsilon_{L}$ in the former replacing $\epsilon$ in the latter. The leading effect of quantum gravity in our quantum field is rather small: it only changes the critical dimension where the theory is renormalizable, manifesting itself in the perturbative expansion parameter and "detaches" the $m_{1}$ subspace from the others. (This was already suggested in Ref. \cite{SCL}.) The application of this operation to two- and three-loop graphs is identical to that in $\phi^{4}$ theory \cite{LL}.
\par We apply the partial-$p$ operation first on $I_{3}$. We find the following expression
\begin{subequations}\label{8}
\begin{eqnarray}
&I_{3}(p_{(n)},\mu_{n}^{*}) = -\frac{1}{(d_{eff}-3)} \Bigl[3 \mu_{n}^ {* 2n} A(p_{(n)},\mu_{n}^{*})\nonumber\\
&+ B(p_{(n)},\mu_{n}^{*}) \Bigr], \label{8a}\\
&A(p_{(n)},\mu_{n}^{*}) = \int  \frac{\Pi_{n=1}^{L} d^{m_{n}} q_{1(n)} d^{m_{n}} q_{2(n)}}{[(\sum_{n=1}^{L} q_{1(n)}^{2n}) + \mu_{n}^{* 2n}]^{2}[(\sum_{n=1}^{L} q_{2(n)}^{2n}) + \mu_{n}^{* 2n})}\nonumber\\
&\frac{1}{[(\sum_{n=1}^{L}(q_{1(n)} + q_{2(n)} + p_{(n)})^{2n}) + \mu_{n}^{* 2n}]}, \label{8b}\\
&B(p_{(n)},\mu_{n}^{*}) = \int  \frac{\Pi_{n=1}^{L} d^{m_{n}} q_{1(n)} d^{m_{n}} q_{2(n)}  p_{(n)}. (q_{1(n)} +q_{2(n)} +p_{n})}{[(\sum_{n=1}^{L} q_{1(n)}^{2n}) + \mu_{n}^{* 2n}] 
[(\sum_{n=1}^{L} q_{2(n)}^{2n}) + \mu_{n}^{* 2n}]}\nonumber\\
&\frac{1}{[(\sum_{n=1}^{L}(q_{1(n)} + q_{2(n)} + p_{(n)})^{2n}) + \mu_{n}^{* 2n}]^{2}}.\label{8c}
\end{eqnarray}
\end{subequations}
\par One subdiagram of the  integral $A_{n}$ can be solved using the orthogonal approximation \cite{L2,CL2}. By integrating first over the quadratic momenta and using another Feynman parameter, we get to
\begin{eqnarray}\label{9}
& A(p_{(n)},\mu_{n}^{*}) = \frac{1}{2} (1-\frac{\epsilon_{L}}{2} \psi(2 -\sum_{n=2}^{L} \frac{m_{n}}{2n})) \Gamma(2+\frac{\epsilon_{L}}{2})\nonumber\\
& \int_{0}^{1} dx [x(1-x)]^{-\frac{\epsilon_{L}}{2}}  \int_{0}^{1} dy y^{\frac{\epsilon_{L}}{2} -1} (1-y) \int \Pi_{n=1}^{L} d^{m_{n}} q_{1(n)} \nonumber\\
&\;\; \times \Bigl[q_{1(1)}^{2} + 2q_{1(1)}.p_{(1)}y + p_{(1)}^{2} y + \sum_{n=2}^{L} (q_{1(n)} + 
p_{(n)})^{2n} y\nonumber\\
& + \mu_{n}^{* 2n}[1-y+ \frac{y}{x(1-x)}]\Bigr]^{-2 -\frac{\epsilon_{L}}{2}}.
\end{eqnarray}
Now we apply the parametric dissociation transform ($PDT$), namely we get rid of all the 
external momenta in the $A(p_{(n)}, \mu_{n}^{*})$ integral by setting $y=0$ in the integrand of the last loop integral \cite{LL}. We then find $A(p_{(n)},\mu_{n}^{*} )_{PDT}= \frac{\mu_{n}^{ * -2n \epsilon_{L}}}{2 \epsilon_{L}^{2}}\Bigl[1 + \epsilon_{L}(2h_{m_{L}} -\frac{3}{2}) 
+ \epsilon_{L}^{2}(2h_{m_{L}}^{2} - 3h_{m_{L}} + \frac{7}{4} + \frac{1}{4}[\psi^{'}(2 - \sum_{n=2}^{L} \frac{m_{n}}{2n}) + \psi^{'}(1)] )\Bigr]$. As it is going to be shown in a moment, 
the diagrams left in the two-point vertex part must be subtracted from their values at zero external momenta. Employing the $PDT$ in the zero momenta analogue $A(p_{(n)}=0, \mu_{n}^{*})$ we find 
$(A(p_{(n)}, \mu_{n}^{*}) - A(0, \mu_{n}^{*}))_{PDT}=0$. The remaining integral can be shown to be given by $B(p_{(n)},\mu_{n}^{*}) = 
\frac{\mu_{n}^{* -2n \epsilon_{L}} \sum_{n=1}^{L} p_{(n)}^{2n}}{8 \epsilon_{L}}
\Bigl[ 1 + (2h_{m_{L}} -\frac{7}{4})\epsilon_{L} - 2 \epsilon_{L} L_{3}(p_{n)}, \mu_{n}^{*})\Bigr]$, where 
$L_{3}(p_{(n)},\mu_{n}^{*})= \int_{0}^{1} dx \int_{0}^{1} dy (1-y) ln\Bigl[y(1-y)\frac{\sum_{n=1}^{L}p_{(n)}^{2n}}{\mu_{n}^{* 2n}} + 1 -y + \frac{y}{x(1-x)}\Bigr]$.
\par Altogether, we find the result $(I_{3}(p_{(n)}, \mu_{n}^{*}) - I_{3}(0, \mu_{n}^{*}))_{PDT} = -\frac{\mu_{n}^{* -2n \epsilon_{L}} \sum_{n=1}^{L} p_{(n)}^{2n}}{8 \epsilon_{L}}\Bigl[1+(2h_{m_{L}} -\frac{3}{4})\epsilon_{L} - 2 \epsilon_{L} L_{3}(p_{(n)},\mu_{n}^{*})\Bigr]$.
\par We can define the {\it restriction} of all integrals to a 
given mass/momentum subspace by writing  
\begin{eqnarray}\label{10}
&(I_{3n}(p_{(n)}, \mu_{n}) - I_{3n}(0, \mu_{n}))_{PDT} = -\frac{\mu_{n}^{-2n \epsilon_{L}} p_{(n)}^{2n}}{8 \epsilon_{L}}\Bigl[1+\nonumber\\
&(2h_{m_{L}} -\frac{3}{4})\epsilon_{L} - 2 \epsilon_{L} \tilde{L}_{3}(p_{(n)},\mu_{n})\Bigr],
\end{eqnarray}
with  $\tilde{L}_{3}(p_{(n)},\mu_{n})= \int_{0}^{1} dx \int_{0}^{1} dy (1-y) ln\Bigl[y(1-y)\frac{p_{(n)}^{2n}}{\mu_{n}^{2n}} + 1 -y + \frac{y}{x(1-x)}\Bigr]$. A similar restriction for the four-point and composite vertex part is to work with the restricted integral  $\tilde{L}(P(n),\mu_{n}) = \int_{0}^{1} dx ln\Bigl[\frac{P_{(n)}^{2n}}{\mu_{n}^{2n}} x(1-x) + 1 \Bigr]$. We will use the restriction definition extensively in our diagrammatic expansion, since it has no effect in the $PDT$. 
\par We apply now the partial-$p$ operator in the integral $I_{5}(p_{(n)}, \mu_{n}^{*})$. We find
\begin{subequations}\label{11}
\begin{eqnarray}
& I_{5}(p_{(n)},\mu_{n}^{*}) = -\frac{2}{(3d_{eff}-10)} \Bigl[\mu_{n}^{* 2n} (C_{1}(p_{(n)},\mu_{n}^{*}) \nonumber\\
&+ 4 C_{2}(p_{(n)},\mu_{n}^{*})) + 2 D(p_{(n)},\mu_{n})\Bigr],\label{11a}\\
&C_{1}(p_{(n)},\mu_{n}^{ *}) = \int \frac{\Pi_{n=1}^{L} d^{m_{n}} q_{1(n)} d^{m_{n}} q_{2(n)} 
d^ {m_{n}} q_{3(n)}}{[(\sum_{n=1}^{L}q_{1(n)}^{2n}) + \mu_{n}^{* 2n}]^{2} [(\sum_{n=1}^{L}q_{2(n)}^{2n}) + \mu_{n}^{* 2n}]}\nonumber\\
&\frac{1}{[(\sum_{n=1}^{L} q_{3(n)}^{2n}) + \mu_{n}^{* 2n}][(\sum_{n=1}^{L}(q_{1(n)} + q_{2(n)} + p_{(n)})^{2n}) + \mu_{n}^{* 2n}]}\nonumber\\
& \times \; \frac{1}{[(\sum_{n=1}^{L}(q_{1(n)} + q_{3(n)} + p_{(n)})^{2n}) + \mu_{n}^{* 2n}]},\label{11b}\\
&C_{2}(p_{(n)},\mu_{n}^{ *}) = \int \frac{\Pi_{n=1}^{L} d^{m_{n}} q_{1(n)} d^{m_{n}} q_{2(n)} 
d^ {m_{n}} q_{3(n)}}{[(\sum_{n=1}^{L} q_{1(n)}^{2n}) + \mu_{n}^{* 2n}][(\sum_{n=1}^{L} q_{2(n)}^{2n}) + \mu_{n}^{* 2n}]^{2}}\nonumber\\
&\frac{1}{[(\sum_{n=1}^{L} q_{3(n)}^{2n}) + \mu_{n}^{* 2n}][(\sum_{n=1}^{L}(q_{1(n)} + q_{2(n)} + p_{(n)})^{2n}) + \mu_{n}^{* 2n}]}\nonumber\\
& \times \; \frac{1}{[(\sum_{n=1}^{L}(q_{1(n)} + q_{3(n)} + p_{(n)})^{2n}) + \mu_{n}^{* 2n}]},\label{11c}\\
& D(p_{(n)},\mu_{n}^{*}) =  \int  \frac{\Pi_{n=1}^{L} d^{m_{n}} q_{1(n)} d^{m_{n}} q_{2(n)} d^{m_{n}} q_{3(n)}}{[(\sum_{n=1}^{L} q_{1(n)}^{2n}) + \mu_{n}^{* 2n}][(\sum_{n=1}^{L} q_{2(n)}^{2n}) + \mu_{n}^{* 2n}]}\nonumber\\
& \frac{p_{(n)}.(q_{1(n)} + q_{2(n)} + p_{(n)})}{[(\sum_{n=1}^{L} q_{3(n)}^{2n}) + \mu_{n}^{* 2n}][(\sum_{n=1}^{L}(q_{1(n)} + q_{2(n)} + p_{(n)})^{2n}) + \mu_{n}^{* 2n}]^{2}}\nonumber\\
& \frac{1}{[(\sum_{n=1}^{L}(q_{1(n)} + q_{3(n)} + p_{(n)})^{2n}) + \mu_{n}^{* 2n}]}.\label{11d}
\end{eqnarray}
\end{subequations}
\par By applying the $PDT$ with the same set of parameters appropriate 
to the three-loop order integrals $C_{i}(p_{(n)},\mu_{n}^{*})$, we find 
$C_{1}(p_{(n)},\mu_{n}^{*}) + 4 C_{2}(p_{(n)},\mu_{n}^{*}) = -\frac{\mu_{n}^{* -3n\epsilon_{L}}}{3 \epsilon_{L}^{3}} \Bigl[1 + (3h_{m_{L}} -4)\epsilon_{L} + \frac{\epsilon_{L}^{2}}{2}[13h_{m_{L}}^{2} 
-32h_{m_{L}} + 25 + \frac{3}{4} \psi^{'}(2-\sum_{n=2}^{L} \frac{m_{n}}{2n}) 
- \frac{1}{4} \psi^{'}(1)]\Bigr]$. When this contribution is subtracted from its value at $p_{(n)}=0$ using $PDT$, it vanishes. The integral  $D(p_{(n)},\mu_{n}^{*})$ can be computed and what is left from the original integral $I_{5}$ after $PDT$. Therefore, using the restriction aforementioned we obtain the result
\begin{eqnarray}\label{12}
&(I_{5n}(p_{(n)},\mu_{n}) - I_{5n}(0,\mu_{n}))_{PDT}= -\frac{\mu_{n}^{ -3 n \epsilon_{L}}} {6 \epsilon_{L}^{2}}p_{(n)}^{2n} \nonumber\\
&\Bigl[1 + 3 \epsilon_{L} (h_{m_{L}} - 1) - 3 \epsilon_{L} \tilde{L}_{3}(p_{(n)}, \mu_{n})\Bigr]
\end{eqnarray} 
\par Next, we define the $1PI$ primitively divergent vertex parts in each subspace. Define the three-loop bare mass in the $m_{n}$ subspace by writing 
$\mu_{n}^{2n}= \Gamma_{n}^{(2)} (p_{(n)}=0, \mu_{0n},\lambda_{0n}, \Lambda_{n})$, which can be inverted to express all the vertex parts in terms of $\mu_{n}$ in the diagrammatic expansion and getting rid, at the same time, of all tadpole insertions in the vertex parts which can be multiplicatively renormalized (see Ref. \cite{CL} for the pure $\phi^{4}$ theory).  We shall need only the restriction of the results above to the subspace $m_{n}$. From now on we drop the subscript $PDT$.
\par The bare primitively vertex parts after this maneuver can be written as 
\begin{subequations}\label{13}
\begin{eqnarray}
&&\Gamma_{(n)}^{(2)} (p,\mu_{n},\lambda_{0n}, \Lambda_{N}) =  p_{(n)}^{2n} + \mu_{n}^{2n} - \frac{\lambda_{0n}^{2}(N+2)}{18} \times \nonumber\\
&&[I_{3n}(p_{(n)},\mu_{n}) - I_{3n}(0,\mu_{n})]  \nonumber\\
&& + \frac{\lambda_{0n}^{3} (N+2)(N+8)}{108}[I_{5n}(p_{(n)},\mu_{n})- I_{5n}(0,\mu_{(n)})], \label{13a}\\
&& \Gamma_{(n)}^{(4)} (p_{i (n)},\mu_{n},\lambda_{0 n}) = \lambda_{0 n} \nonumber\\
&&- \frac{\lambda_{0 n}^{2} (N+8)}{18} [I_{2n}(p_{1 (n)} + p_{2 (n)}, \mu_{n})  + 2perms. ] \nonumber\\
&&+ \frac{\lambda_{0 n}^{3} (N^{2} + 6N + 20)}{108} [ I_{2n}^{2} (p_{1(n)} + p_{2(n)},\mu_{n})+ 2 perms. ]\nonumber\\ 
&&+ \frac{\lambda_{0 n}^{3} (5N+22)}{54}[ I_{4n} (p_{i(n)}, \mu_{n}) + 5 perms. ],\label{13b}\\
&&\Gamma_{(n)}^{(2,1)} (p_{1(n)}, p_{2(n)}; Q_{(n)}, \mu_{n},\lambda_{0 n}) = 1 -  \frac{\lambda_{0 n}(N+2)}{18} \times \nonumber\\
&&[I_{2n}(p_{1(n)} + p_{2(n)}, \mu_{n}) + 2perms.] +  \frac{\lambda_{0 n}^{2}(N+2)^{2}}{108}\times \nonumber\\
&&[I_{2n}^{2} (p_{1(n)}+p_{2(n)}, \mu_{n})  + 2 perms. ] +  \frac{\lambda_{0 n}^{2}(N+2)}{36} \times \nonumber\\
&&[ I_{4n}(p_{1 (n)}, p_{2(n)}, Q_{(n)}, \mu_{n}) + 5 perms. ].\label{13c}
\end{eqnarray}
\end{subequations}
\par The definitions $\lambda_{0n} =u_{0n} \mu_{n}^{n \epsilon}$ and $g_{n}=u_{n}\mu_{n}^{n \epsilon}$ in terms of the dimensionless bare ($u_{0n}$) and renormalized ($u_{n}$) coupling constants will be useful in what follows. The dependence of the integrals in the overall mass scales $\mu_{n}$ is offset by the contribution of the bare (and renormalized) coupling constants. This occurs not only in the perturbative primitively divergent vertex parts, but also in all vertex parts that can be renormalized multiplicatively, resulting in a perturbative expansion only in the dimensionless coupling constants. 
\par Multiplicative renormalizability is the statement that if the primitively divergent vertex are renormalized multiplicatively in a given loop order, then all multiplicatively renormalized vertex 
part with arbitrary number of composite fields are related to the bare ones through
$\Gamma_{R(n)}^{(N,L)}(p_{i(n)};Q_{i(n)}; \frak{m}_{n}, u_{n})= Z_{\phi (n)}^{\frac{N}{2}} 
Z_{\phi^{2} (n)}^{L} \Gamma_{(n)}^{(N,L)}(p_{i(n)}; Q_{i(n)}; \mu_{n}, u_{0n})$, where 
$\frak{m}_{n}$ are the renormalized masses. First we renormalize the primitively divergent parts utilizing this definition. We demand that 
\begin{subequations}\label{14}
\begin{eqnarray}
&&\Gamma_{R(n)}^{(2)}(p_{(n)},\frak{m}_{n},u_{n}) = Z_{\phi (n)} \Gamma_{(n)}^{(2)}(p_{(n)},\mu_{n}, u_{0n}),\label{14a}\\
&&\Gamma_{R(n)}^{(4)}(p_{i(n)},\frak{m}_{n},u_{n}) = Z_{\phi (n)}^{2} \Gamma_{(n)}^{(4)}(p_{i(n)},\mu_{n},u_{0n}),\label{14b}\\
&&\Gamma_{R(n)}^{(2,1)}(p_{1(n)},p_{2(n)}; Q_{(n)},\frak{m}_{n},u_{n}) = \bar{Z}_{\phi^{2} (n)}\times\nonumber\\
&& \Gamma_{(n)}^{(2,1)}(p_{1(n)},p_{2(n)};Q_{(n)},\mu_{n},u_{0n}),\label{14c}
\end{eqnarray}
\end{subequations}
are finite by minimal subtraction ($MS$) of the dimensional poles 
($\bar{Z}_{\phi^{2} (n)} \equiv Z_{\phi (n)} Z_{\phi^{2} (n)}$). Then, we perform the expansion of these functions in terms of the dimensionless couplings $u_{n}$ as  
\begin{subequations}\label{15}
\begin{eqnarray} 
&& u_{0n} = u_{n}(1+a_{1n} u_{n} +a_{2n} u_{n}^{2}), \label{15a} \\
&& Z_{\phi (n)} = 1+b_{2n} u_{n}^{2} +b_{3n} u_{n}^{3}, \label{15b} \\
&& \bar{Z}_{\phi^{2} (n)} = 1 + c_{1n} u_{n} +c_{2n} u_{n}^{3}. \label{15c} 
\end{eqnarray}
\end{subequations}
\par By defining the three-loop renormalized masses 
$\frak{m}_{n}^{2n}=Z_{\phi (n)} \mu_{n}^{2n}$ we can find all coefficients using the diagrammatic expansion: $b_{2n}$ from the singular part 
of the two-loop contribution of $\Gamma_{R(n)}^{(2)}$and  $a_{1n}$ ($c_{1n}$) from the one-loop contribution from $\Gamma_{R(n)}^{(4)}$ ($\Gamma_{R(n)}^{(2,1}$). Now $b_{3n}$ can be determined and proves that the coefficient of the integrals $\tilde{L}_{3}(p_{(n)},\mu_{n})$ vanishes; $a_{2n}$ and $c_{2n}$ are then determined with the cancellation of the coefficient of the integrals  $\tilde{L}(p_{(n)},\mu_{n})$.  We then find:
\begin{subequations}\label{16}
\begin{eqnarray}
&u_{0n} = u_{n}\Bigl(1+ \frac{(N+8)}{6 \epsilon_{L}} u_{n} +\Bigl[\frac{(N+8)^{2}}
{36 \epsilon_{L}^{2}} -\frac{(3N+14)}{24 \epsilon_{L}}\Bigr]\nonumber\\
&u_{n}^{2}\Bigr),\label{16a}\\
&\bar{Z}_{\phi^{2} (n)}= 1+ \frac{(N+2)}{6 \epsilon_{L}} u_{n} + \Bigl[\frac{(N+2)(N+5)}{36 \epsilon_{L}^{2}} - \frac{(N+2)}{24 \epsilon_{L}}\Bigr]\nonumber\\
& \times u_{n}^{2},\label{16b} \\ 
& Z_{\phi(n)} = 1 -\frac{(N+2)}{144 \epsilon_{L}} u_{n}^{2} - \Bigl[\frac{(N+2)(N+8)}{1296 \epsilon_{L}^{2}} + \frac{N+2)(N+8)}{5184 \epsilon_{L}}\Bigr]\nonumber\\
&\times u_{n}^{3}. \label{16c} 
\end{eqnarray}
\end{subequations}
\par  The method can employed in the  calculation of critical exponents using the Callan-Symanzik  ($CS$) method \cite{CS}, since in the Lagrangian (1), the mass $\frak{m}_{n}$ is proportional to 
$|T-T_{C}|^{\frac{1}{n}}$, with $\frak{m}_{n} \neq 0$ but still in the critical region.  By applying the operator $\frak{m}_{n} \frac{\partial}{\partial {\frak{m}_{n}}}$ on $\Gamma_{R}^{(N,L)}$, 
we find $\Bigl[\frak{m}_{n} \frac{\partial}{\partial \frak{m}_{n}} + \beta_{n}(u_{n}) \frac{\partial}{\partial u_{n}} -\frac{N}{2} \gamma_{\phi(n)}(u_{n}) + L \gamma_{\phi^{2(n)}}(u_{n})\Bigr] \Gamma_{R(n)}^{(N,L)}(p_{i};Q_{j},\frak{m}_{n},u_{n})= \frac{(2-\gamma_{\phi (n)}(u_{n}))\frak{m}_{n}^{2n}}{Z_{\phi (n)}Z_{\phi^{2} (n)}}  \Gamma_{R(n)}^{(N,L+1)}(p_{i};Q_{j},0,\frak{m}_{n},u_{n})$, without using normalization conditions ($NC$) for $N=2,L=0$ as this equation is usually obtained \cite{CL2,amit}. In our method, $\frak{m}_{n}$ is defined for arbitrary external momenta. The counterpart of this statement in the $CS$ equation is to set $\bar{Z}_{\phi^{2} (n)}=1$ (tree-level value). This is the condition for the "covariance" of the $CS$ equation either using $NC$ or $MS$ and we find 
\begin{eqnarray}\label{17}
&&\Bigl[\frak{m}_{n} \frac{\partial}{\partial \frak{m}_{n}} + \beta_{n}(u_{n}) \frac{\partial}{\partial u_{n}} -\frac{N}{2} \gamma_{\phi (n)}(u_{n}) \nonumber\\
&&+ L \gamma_{\phi^{2(n)}}(u_{n})\Bigr] \Gamma_{R(n)}^{(N,L)}(p_{i};Q_{j},\frak{m}_{n},u_{n})= (2-\gamma_{\phi (n)}(u_{n}))\nonumber\\
&&\frak{m}_{n}^{2n}  \Gamma_{R(n)}^{(N,L+1)}(p_{i};Q_{j},0,\frak{m}_{n},u_{n}).
\end{eqnarray} 
\par In the UV regime, the (rhs) of this equation is neglected in comparison with the (lhs), validating scaling  theory \cite{amit,Wein}.  
\par The Wilson functions 
\begin{eqnarray}\label{18}
&&\beta_{n}(u_{n}) = - n \epsilon_{L} \left(\frac{\partial lnu_{0}}{\partial u}\right)^{-1} = n u_{n} 
\Bigl[ -\epsilon_{L} + \frac{(N+8)}{6} \nonumber\\
&&u_{n} -\frac{(3N+14)}{12} u_{n}^{2} \Bigr],
\end{eqnarray}
has a nontrivial (repulsive) $UV$ fixed point ($\beta_{n} (u_{n \infty})=0$) at $u_{n \infty} = \frac{6 \epsilon_{L}}{N+8)}\Bigl[1 + \frac{3(3N+14)}{(N+8)^{2}} \epsilon_{L} \Bigr]$. The functions
$\gamma_{\phi (n)}(u_{n})= \Bigl[\beta_{n}(u_{n}) \frac{\partial ln Z_{\phi(n)}}{\partial u_{n}}\Bigr] = n\Bigl[ \frac{(N+2)}{72} u_{n}^{2} - \frac{(N+2)(N+8)}{1728} u_{n}^{3}\Bigr]$ computed at the fixed point yields $\eta_{n}$ up to three-loop order. It is given by
\begin{equation}\label{19}
\eta_{n} = n \frac{(N+2)\epsilon_{L}^{2}}{2(N+8)^{2}}\Bigl[1 + \Bigl(\frac{6(3N+14)}{(N+8)^{2}} -\frac{1}{4}\Bigr)\epsilon_{L} \Bigr]
\end{equation} 
The functions $\bar{\gamma}_{\phi^{2}(n)} (u_{n})= -\Bigl[ \beta_{n}(u_{n}) \frac{\partial ln \bar{Z}_{\phi^{2(n)}}}{\partial u_{n}}\Bigr]= n u_{n} \frac{(N+2)}{6} [1 - \frac{u_{n}}{2}]$  computed at the fixed point together with the identity $\nu_{n}^{-1}= 2n - \bar{\gamma}_{\phi^{2}(n)} (u_{\infty (n)}) - \eta_{n}$ lead to the correlation length exponents 
\begin{equation}\label{20}
\nu_{n}= \frac{1}{2n} + \frac{(N+2)\epsilon_{L}}{4n(N+8)} + \frac{(N+2)(N^{2} + 23N + 60)\epsilon_{L}^{2}}{8n(N+8)^{2}}.
\end{equation}
A detailed discussion will be reported in the near future.
\par When the $m_{1}$ subspace is extended to Minkowski spacetime, the massive quantum scalar fields in  Lifshitz spacetimes interact with the competing directions through the leading order quantum gravity effect. The subspace $m_{1}<4$ can have its  diffeomorphism invariance restored in $S_{QFT}$ so that general relativity can be formulated on it as explained. Thus an observer living on the $m_{1}$ subspace 
would feel a gravitational field weakly disturbed by the extra dimensions, having the impression that they are not there. The other terms with  higher momentum powers in the "extra" directions break Lorentz invariance only in the "bulk" ($d$ dimensions). The exponents corresponding to the power of momenta in each subspace are fixed. The only way to come back to the original $\phi^{4}$ is to set $m_{n}=0$, what corresponds to a massive quantum scalar field in a classical Minkowski/$GR$ ($d=4$) background. 
\par To summarize, we have shown for the first time the equivalence of the $CECI$ model with 
a scalar quantum field theory in a Lifshitz spacetime. Even its flat version already includes the first corrections of quantum gravity emerging from the competition subspaces. In addition, the quantum theory of the massive scalar field through the usage of the partial-$p$ operation along with the $PDT$, etc., results in a masslesslike massive method of minimal subtraction. It is quite simple, although different from the massless one \cite{L2,ML}. We computed critical exponents as an application of the method. Massive nonlinear sigma models with competing interactions  studied using the $\frac{1}{N}$ expansion \cite{GBG} could  be complemented using the present minimal subtraction. 
\par We would like to acknowledge partial support from CAPES (Brazilian agency) through the PROEX Program 534/2018 grant number 23038.003382/2018-39.

\end{document}